\documentclass{ws-mpla}
\newcommand{\bra}[1]{\mbox{$\langle #1|$}}
\newcommand{\ket}[1]{\mbox{$|#1\rangle$}}

\begin{document}

\markboth{A. Salam, T. Mart, and K. Miyagawa}
{Neutral Kaon Photoproduction on the Deuteron}

\catchline{}{}{}{}{}

\title{NEUTRAL KAON PHOTOPRODUCTION\\ ON THE DEUTERON}

\author{\footnotesize A. SALAM and T. MART}

\address{Departemen Fisika, FMIPA, Universitas Indonesia\\ Depok 16424,
  Indonesia}

\author{K. MIYAGAWA}

\address{Simulation Science Center, Okayama University of Science\\
1-1 Ridai-cho, Okayama 700-0005,Japan
}

\maketitle

\pub{Received (Day Month Year)}{Revised (Day Month Year)}

\begin{abstract}
Neutral kaon photoproduction on the deuteron has been investigated by
including the final state effects and compared with the experimental data. 
Comparison shows that the models used in this calculation 
can reproduce the data in the $\Sigma$ channel regions fairly well 
but still give over predictions in the $\Lambda$ channel.
It seems that the tensor target asymmetries are more suitable for studying 
the final state effects. 
The extractions of the elementary photoproduction amplitude are also 
demonstrated.

\end{abstract}

\ccode{PACS Nos.: include PACS Nos.}

\section{Introduction}	

The mostly investigated kaon photoproduction on the nucleon are the proton
channels~\cite{DaF96,MaB00}, i.e., $\gamma p\rightarrow K^+\Lambda$ and
$\gamma p\rightarrow K^+\Sigma^0$, since the experimental data are available 
for these channels~\cite{Boc94,Tra98,Goers:1999sw}. Unfortunately, it is not
the case for the neutron channels, since free neutron targets are not
available. Instead, one can use deuteron as effective neutron targets because
it has a small binding energy and simple structures. Therefore, kaon
photoproduction on the deuteron is the natural candidate for investigating the
kaon photoproduction on the neutron. 

Furthermore, since the reaction on the deuteron yields several particles in
the final state, it is important to include final state effects in the
calculations. It seems that kaon photoproduction on the deuteron can serve
as an alternative reaction for studying the $YN$ interaction through the $YN$
rescattering at the final state. A previous study~\cite{YaM99} has investigated
$YN$ final state interaction effect by using realistic $YN$
potentials. Another previous study~\cite{SaA04} has included $KN$
rescattering effect and $\pi N\rightarrow KY$ intermediate process 
in its calculations 
for charged kaon channels by using separable interactions. 
Concerning the study of neutron channels, Li~{\it et al.}~\cite{LiW92} has
shown the possibility to extract the elementary amplitude from the reaction on
the deuteron.

Very recently an experiment of neutral kaon photoproduction on the deuterium
has been done at the Laboratory of Nuclear Science (LNS), in
Sendai~\cite{Tsu08}. They measured the cross section of the $d(\gamma,K^0)YN$
process at a photon energy around the threshold with forward kaon angles. 

This paper is structured as follow. The formalism for calculating the
transition matrix and observables are shown in Sect.~\ref{formulations}. The
results and comparisons with experimental data are presented in
Sect.~\ref{results} and we close the paper with conclusions in
Sect.~\ref{conclusions}. 

\section{Formulations}\label{formulations}

\begin{figure}[ph]
\centerline{\psfig{file=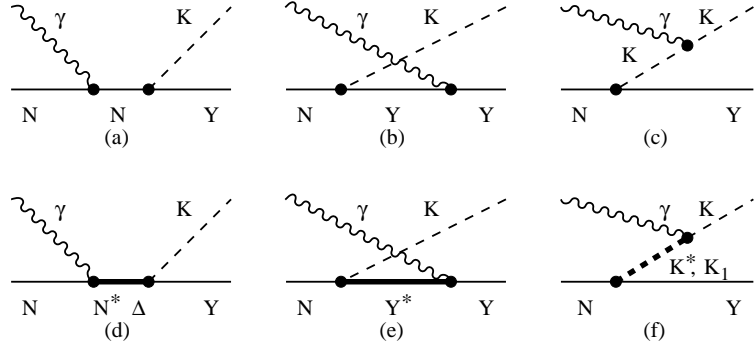,width=.8\textwidth}}
\vspace*{8pt}
\caption{Diagram (a),(b), and (c) : Born terms. Diagram (d), (e), and (f) :
  Resonance terms. Coupling constants are determined by fitting to the
  experimental data   
\label{fig-gnky-drivingterm}}
\end{figure}

In this work we use the KAON-MAID model~\cite{MaB00}, which includes 
the $D_{13}$(1895) resonance beside the Born terms and other resonances as
shown in Figure~\ref{fig-gnky-drivingterm}, for the elementary operator. 
Separate hadronic form factors for each vertex were used in order to restore
gauge invariance. 

Through out the paper we work in the deuteron rest frame. 
For the inclusive process $d(\gamma,K)YN$ the cross section 
is given by
\begin{eqnarray}
\frac{d\sigma}{dp_{K}d\Omega_{K}} &=&  
\int d\Omega^{\,\rm cm}_{Y}\, 
\frac{m_{Y}m_{N}\vert\vec p_{K}\vert^2\vert\vec p^{\,\rm cm}_{Y}\vert}
{4(2\pi)^2E_{\gamma}E_{K}W}\,
\nonumber\\ &&\times\,
\frac{1}{6} \sum_{\mu_{Y}\mu_{N}\mu_{d}\lambda} 
\big\vert\sqrt{2}
\langle\vec p_{Y}\vec p_{N}\mu_{Y}\mu_{N}
\vert T^{\gamma K}_\lambda\vert
\Psi_{\mu_{d}}\rangle\,
\big\vert^2\,,
\label{eq-gdkyn-inclusive-cross-section}
\end{eqnarray}
where $W^2=(P_{d}+Q)^2$ and $\vert\vec p^{\,\rm cm}_{Y}\vert$ 
is the hyperon momentum calculated in the center of mass frame of the two
final baryons. 

\begin{figure}[ph]
\centerline{\psfig{file=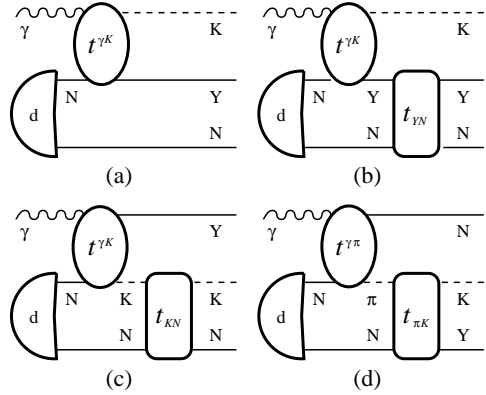,width=.5\textwidth}}
\vspace*{8pt}
\caption{Diagram (a) impulse approximation (IA), (b) $YN$ rescattering, (c)
  $KN$ rescattering, and (d) $\pi N-KY$ process. Total amplitude $T^{\gamma
    K}_\lambda = t^{\gamma K}_{\lambda}+ t^{\gamma K}_{YN}+ t^{\gamma K}_{KN}+
  t^{\gamma K}_{K\pi}$.
\label{fig-gdkyn-amplitudes}}
\end{figure}

The total amplitudes on the deuteron are calculated according to the diagrams
shown in Figure~\ref{fig-gdkyn-amplitudes}. 
For the deuteron wave function and $YN$ interaction we use
the Nijmegen models as in the previous studies~\cite{YaM99,SaA04}. 
Separable interactions rank-1 are applied for the calculations of 
$KN$ rescattering and $\pi N\rightarrow KY$ process. 

With respect to polarization observables, we consider the tensor
target asymmetries $T_{2M}$ which are given by~\cite{Are88}
\begin{eqnarray}
T_{2M} \frac{d\sigma}{d\Omega_{K}} &=&
(2-\delta_{M0})\, {\mathcal Re}\, V_{2M}\,,\quad M=0, 1, 2\,,
\label{eq-gdkyn-T2m}
\end{eqnarray}
where
\begin{eqnarray}
V_{2M} &=& 
\sqrt{15} 
\sum_{\mu_{Y}\mu_{N}\lambda} 
\sum_{\mu^{\prime}_{d}\mu_{d}}
(-1)^{1-\mu^{\prime}_{d}}
\left(\begin{array}{ccc}
    1       &      1                &  2 \\
\mu_{d} & -\mu^{\prime}_{d} & -M 
\end{array}\right) 
\nonumber\\ &&\times\,
\int_{p_K^{\rm min}}^{p_K^{\rm max}} dp_{K} \int d\Omega^{\,\rm cm}_{Y}\,\kappa \,
{\cal M}_{\mu_{Y}\mu_{N}\mu_{d}\lambda}^{*}
{\cal M}_{\mu_{Y}\mu_{N}\mu^{\prime}_{d}\lambda}
\label{eq-gdkyn-V2m}
\end{eqnarray}
with a kinematic factor
\begin{eqnarray}
\kappa &=& 
\frac{m_{Y}m_{N}\vert\vec p_{K}\vert^{2}\vert\vec p^{\,\rm cm}_{Y}\vert}
{24(2\pi)^{2} E_{\gamma}E_{K} W}\,.
\label{eq-gdkyn-kinematic-factor}
\end{eqnarray}
For extracting the elementary amplitudes, we use the expression
\begin{eqnarray}
\frac{d\sigma}{dp_{K}d\Omega_{K}d\Omega_{Y}} &=&
\frac{m_{Y}m_{N}\vert\vec p_{K}\vert^2\vert\vec p_{Y}\vert^2}
{4(2\pi)^2E_{\gamma}E_{K}}\,
\big\vert (E_Y+E_N)\vert\vec p_{Y}\vert 
-E_{Y}\vec Q\cdot\hat p_{Y}\big\vert^{-1}\,
\nonumber\\ &&\times\,
\frac{1}{6}\, D\sum_{\mu_{Y}\mu_{N'}\lambda}
\big\vert\bra{\vec p_{Y}\mu_{Y}} 
t^{\gamma K}_{\lambda} 
\ket{-\vec p_{N}\mu_{N'}}\big\vert^2\,,
\label{eq-gdkyn-exclusive-cross-section-extract}
\end{eqnarray}
where
\begin{eqnarray}
D &=& 
{\textstyle\frac{3}{2}}\vert Y_{00}\vert^2 u_{0}^2 
+\left(
{\textstyle\frac{3}{10}}\vert Y_{20}\vert^2 +
{\textstyle\frac{3}{5}}\vert Y_{21}\vert^2 +
{\textstyle\frac{3}{5}}\vert Y_{22}\vert^2
\right) u_{2}^2\,.
\label{eq-gdkyn-deuteron-factor}
\end{eqnarray}
On the r.h.s. of Eq.~(\ref{eq-gdkyn-exclusive-cross-section-extract}) 
the sum of the squared amplitudes 
of the elementary process $\gamma N \rightarrow KY$ has been completely 
separated from the deuteron wave function.

\section{Results}\label{results}

\begin{figure}[ph]
\hfill
\begin{minipage}[b]{.44\textwidth} 
\centerline{\psfig{file=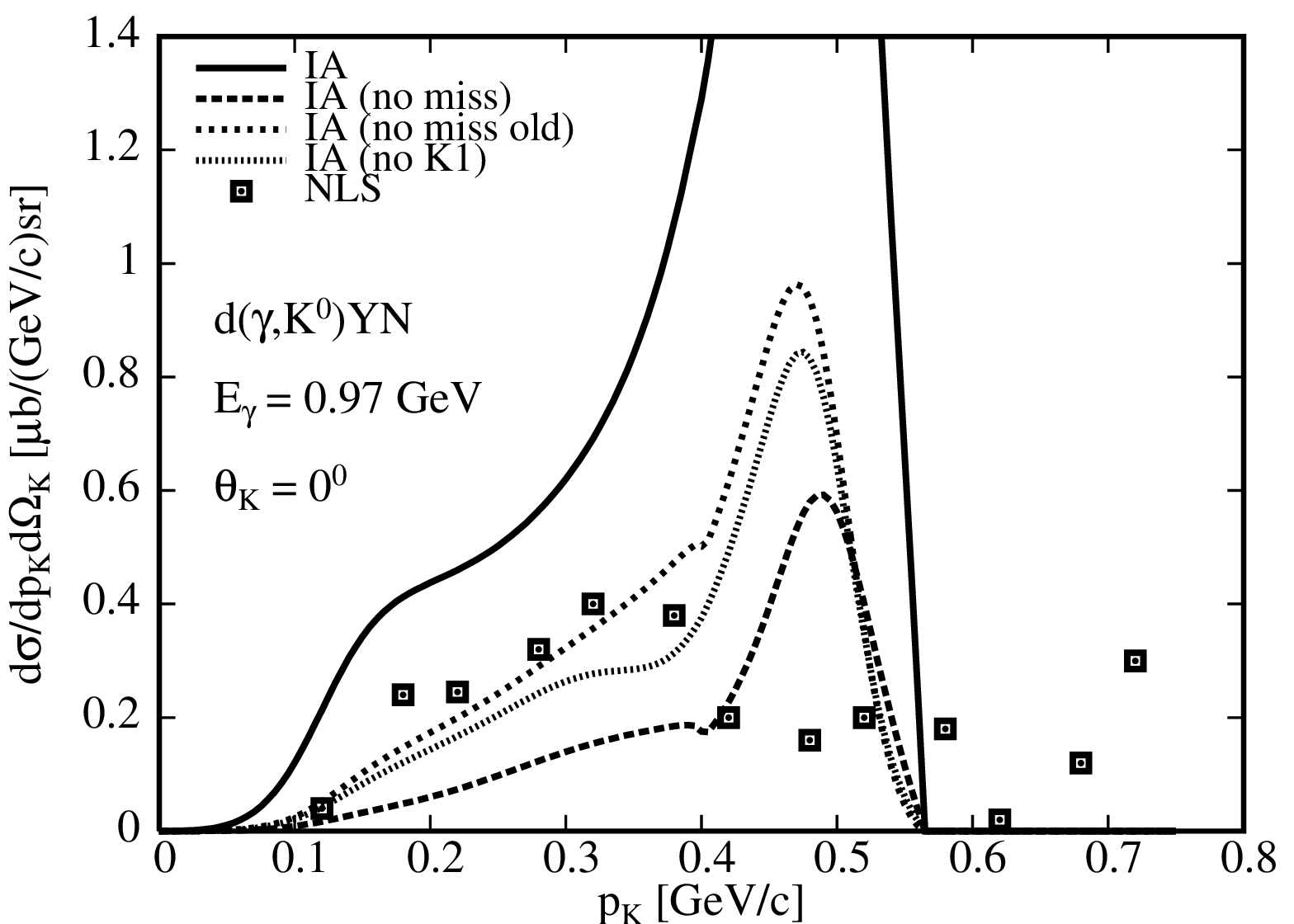,width=1.\textwidth}}
\end{minipage} 
\hfill
\begin{minipage}[b]{.44\textwidth} 
\centerline{\psfig{file=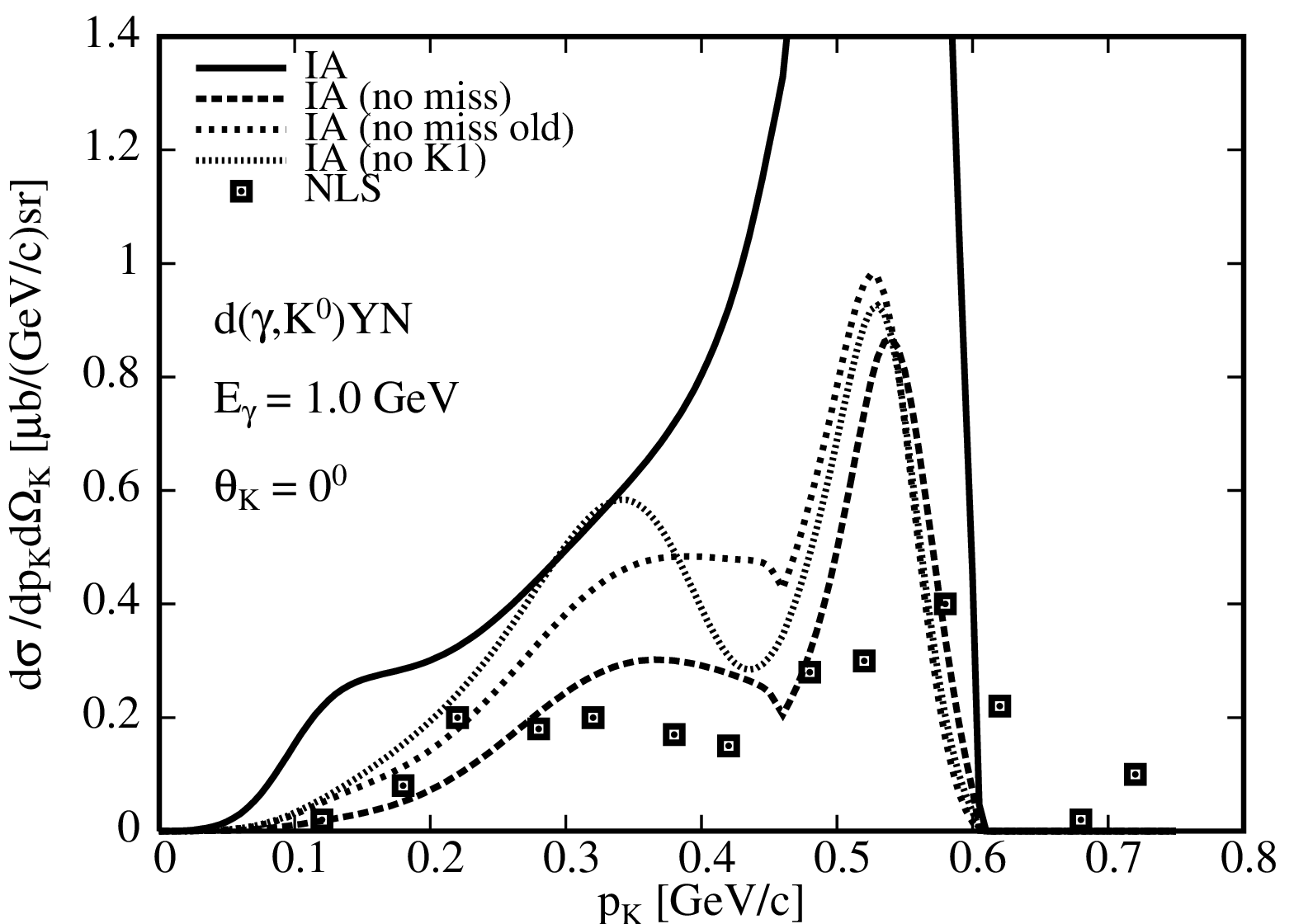,width=1.\textwidth}}
\end{minipage}
\vspace*{8pt}
\caption{Inclusive cross section as a function of kaon momentum $p_K$ 
at photon energy $E_{\gamma}$ = 0.97 GeV (left panel) 
and 1.0 GeV (right panel)
for $d(\gamma,K^0)YN$ in impulse approximation (IA) 
with different sets of KAON-MAID coupling constants. 
Data are taken from Tsukada~{\it et al.}
\label{fig-comparison-experimental-data}}
\end{figure}

Figure~\ref{fig-comparison-experimental-data} shows comparisons between the
inclusive cross sections of the models and of the experimental data 
for reaction $d(\gamma,K^0)KY$ as a function of kaon momentum $p_K$ 
at forward kaon angle $\theta_K=0$. 
The left panel for photon energy $E_{\gamma}=0.97$ GeV 
and the right one for $E_{\gamma}=1.0$ GeV. 
The curves are calculated in the impulse approximation only 
and the data are taken from Tsukada~{\it et al.}~\cite{Tsu08}.
There are several sets of coupling constant by KAON-MAID model. 
The first one is the set obtained after fitting with the experimental data 
by including the $D_{13}$(1895) resonance or the so called missing resonance. 
It is shown by the solid line in the figure and we call this as set A. 
The other ones are the sets without the missing resonance, which are shown by
the different dashed lines in the figure and we call these as sets B. 

As can be seen from the figure, the set A gives over predictions, 
while the sets B can describe the data fairly well in the $\Sigma$ channels
(up to $p_K=0.4$ GeV/c for $E_{\gamma}=0.97$ GeV 
and up to $p_K=0.5$ GeV/c for $E_{\gamma}=1.0$ GeV) 
and are still poor for the $\Lambda$ channel (up to threshold of $p_K$ for
both photon energies). 
The discrepancies between the set A and the experimental data 
are originated from the value of coupling constant 
in the vertex of $\gamma KK1$ of the set A 
(see diagram (f) of Figure~\ref{fig-gdkyn-amplitudes}). 
We can see it from the figure as follow. 
By setting the value of $\gamma KK1$ coupling to zero in the set A 
(indicated with no $K1$ in the figure), then the cross section
reduces to the value in the order of sets B and the experimental data. 

\begin{figure}[ph]
\hfill
\begin{minipage}[b]{.44\textwidth} 
\centerline{\psfig{file=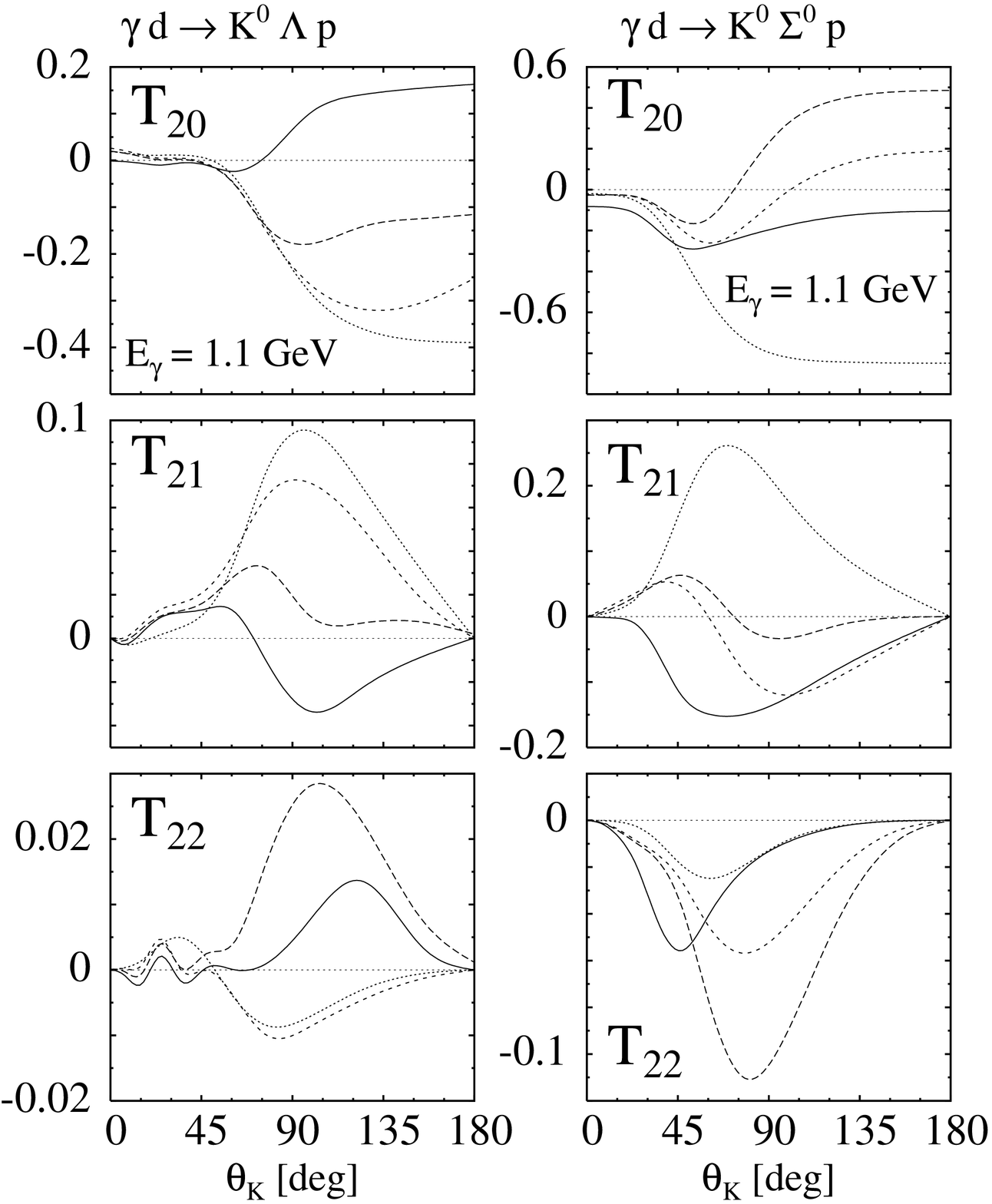,width=1.\textwidth}}
\vspace*{8pt}
\caption{Tensor asymmetries $T_{2M}$
as a function of kaon angle $\theta_{K}$ 
for photon energy $E_{\gamma}$ = 1.1 GeV. 
The left panels for $\gamma d \rightarrow K^{0}\Lambda p$ 
and the right for $\gamma d \rightarrow K^{0}\Sigma^{0} p$.
The solid line for IA+YN+KN+$\pi$N, the dashed for IA+YN+KN, 
the short-dashed for IA+YN, and the dotted for IA.
\label{fig-tensor-asymmetry-1100}}
\end{minipage} 
\hfill
\begin{minipage}[b]{.44\textwidth} 
\centerline{\psfig{file=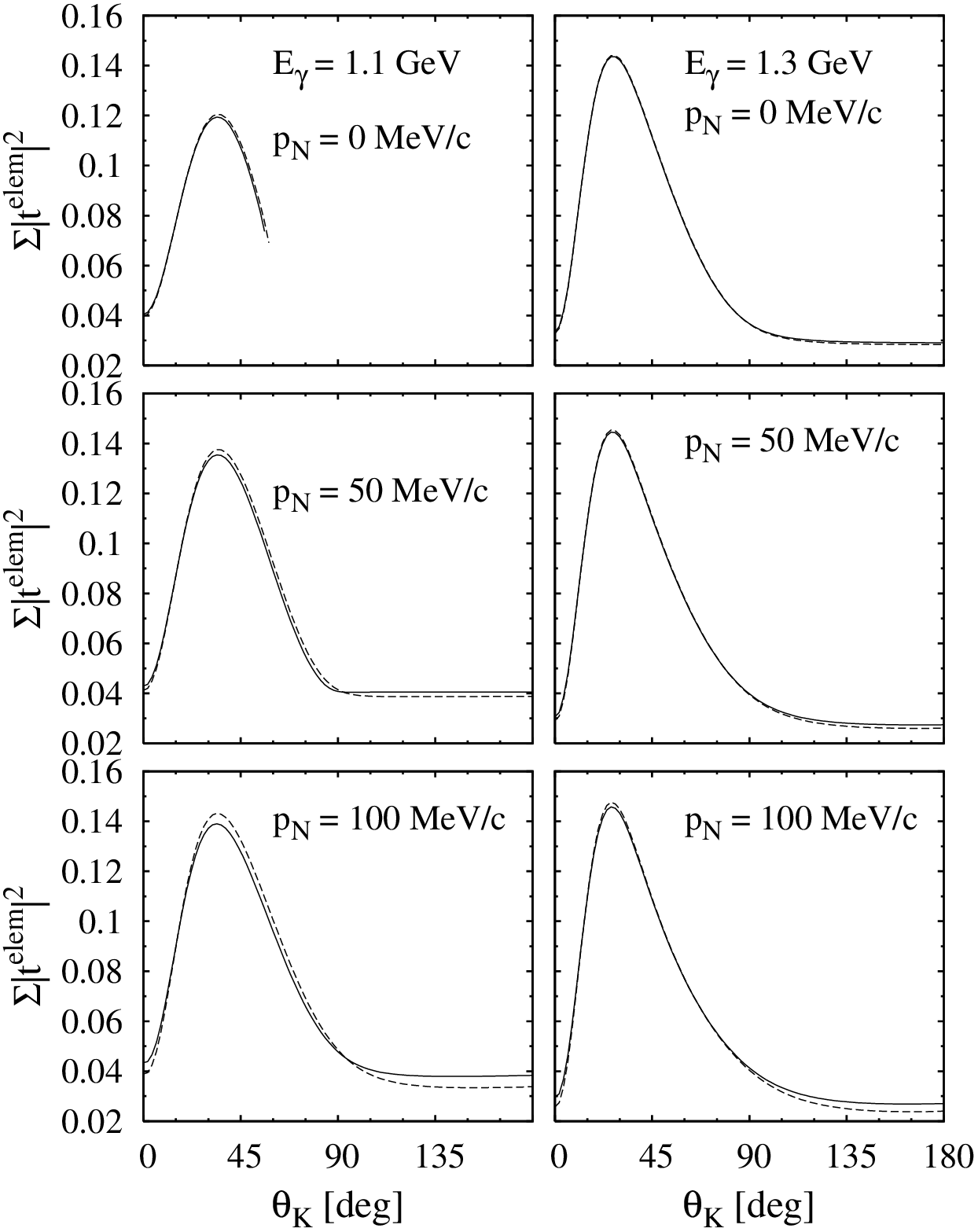,width=.84\textwidth}}
\vspace*{8pt}
\caption{Elementary amplitude as a function of kaon angle $\theta_{K}$ 
for photon energies $E_{\gamma}$ = 1.1 GeV (left panels)
and 1.3 GeV (right panels)
with $\vec p_{N}$ fixed at $\theta_{N} = 30^\circ$
and $\phi_{N} = 180^\circ$, but 
the magnitude varied with 0, 50, 100 MeV/$c$. 
The solid line for the extracted ones
and the dash ones are obtained from the free-process.
\label{fig-extraction-elementary-amplitude}}
\end{minipage}
\end{figure}

Figure~\ref{fig-tensor-asymmetry-1100} shows the tensor target asymmetries 
as a function of kaon angle $\theta_K$ 
for photon energy $E_{\gamma}$ = 1.1 GeV. 
The solid lines are obtained after including all final state effects, the
dashed ones only IA+YN+KN, the short-dashed ones only IA+YN, and the
dotted ones only impulse approximation (IA). 
As can be seen in the figure, the tensor target asymmetries are very sensitive
to the final state effects.

Figure~\ref{fig-extraction-elementary-amplitude} shows the elementary
amplitude as a function of kaon angle $\theta_K$ for different photon energy
and different spectator proton momentum. 
The solid lines are the extracted elementary amplitudes
from the reaction on the deuteron, while the dashed ones are obtained from the
free process. As can be seen in the figure that the agreement between the
extracted and the ones from the free process are very good 
especially in the quasi-free scattering kinematics, 
where the spectator proton has zero momentum.

\section{Conclusions}\label{conclusions}

The sets B of KAON-MAID models describe the experimental data better than the
set A. This indicates  that the data on the deuteron can be used to study
the elementary process. Tensor target asymmetries are more suitable for
investigating the $YN$ interaction through the rescattering effects at the
final state of $\gamma d\rightarrow KYN$. For the process $d(\gamma,K^0Y)N$,
the rescattering effects are negligibe at the quasi-free scattering (QFS)
kinematics. This gives us a possibility to extract the elementary amplitude.
At QFS kinematics the extracted and the free-process amplitudes agree
well. This demonstrates that kaon photoproduction on the deuteron in the QFS
kinematics can be used for investigating the elementary process in the neutron
channels. This suggests that QFS kinematics is suitable for
measurements.  

\section*{Acknowledgments}

Some parts of this work were done during one of the authors (AS) staying at 
Okayama University of Science, Okayama and Tohoku University, Sendai. 
Therefore AS would like to thank the people form both universities for warm
hospitalities. This work was also supported by Fakultas Matematika and Ilmu
Pengetahuan Alam, Universitas Indonesia, Depok.

\end{document}